\documentstyle[preprint,aps]{revtex}
\begin{document}
\draft
\title{Diagrammatic approach in the variational coupled-cluster method}
\author{Y. Xian}
\address{School of Physics and Astronomy, The University of Manchester,
Manchester M13 9PL, UK}
\date{\today }
\maketitle

\begin{abstract}
Recently, as demonstrated by an antiferromagnetic spin-lattice
application, we have successfully extended the coupled-cluster method
(CCM) to a variational formalism in which two sets of distribution
functions are introduced to evaluate Hamiltonian expectation. We
calculated these distribution functions by employing an algebraic
scheme. Here we present an alternative calculation based on a
diagrammatic technique. Similar to the method of correlated-basis
functionals (CBF), a generating functional is introduced and
calculated by a linked-cluster expansion in terms of diagrams which
are categorized and constructed according to a few simple rules and
using correlation coefficients and Pauli exclusion principle (or Pauli
line) as basic elements. Infinite resummations of diagrams can then be
done in a straightforward manner. One such resummation, which includes
all so-called ring diagrams and ignores Pauli exclusion principle,
reproduces spin-wave theory (SWT). Approximations beyond SWT are also
given. Interestingly, one such approximation including all so-called
super-ring diagrams by a resummation of infinite Pauli lines in
additional to resummations of ring diagrams produces a convergent,
precise number for the order-parameter of the one-dimensional
isotropic model, contrast to the well-known divergence of SWT. We also
discuss the direct relation between our variational CCM and CBF and
discuss a possible unification of the two theories.
\end{abstract}


\section{introduction}

A microscopic quantum many-body theory is mainly to study correlations
between the constituent particles of a quantum system. One of the
most successful quantum many-body theories is the method of correlated
basis functionals (CBF) \cite{fee} which includes real-space many-body
correlation functions in the ground state and employs similar
techniques as in classical statistical mechanics to calculate the
corresponding distribution functions. The CBF has proved to be one of
very few many-body theories capable of dealing with strongly
correlated boson systems in liquid phase. Another successful quantum
many-body theory is the coupled-cluster method (CCM) \cite{hhc,ck,ciz}
in which excitation operators with respect to an uncorrelated model
state are employed to construct the many-body correlations in an
exponentiated operator factor in the ket ground-state. The
distribution functions are not needed in the CCM as the Hamiltonian
expectation value is straightforwardly calculated as a finite order
polynomial in terms of the correlation coefficients.  This is because
the bra ground-state in the CCM is not the hermitian conjugate of the
ket ground-state but is in a simple, linear form \cite{arp}. The CCM
has proved to be one of most powerful techniques in calculating
ground-state energy for many non-liquid fermion systems such as atoms,
molecules and electron gas\cite{bb}. However, a general many-body
theory capable of dealing with strongly correlated fermion systems in
liquid phase is still needed.

In our earlier papers \cite{yx1}, by an application to bipartite
quantum antiferromagnetic lattice systems, we have extended the CCM to
a variational formalism where, in contrast to the traditional CCM, the
bra and ket ground-states are hermitian conjugate to one-another. Two
sets of the distribution functions were introduced to evaluate
Hamiltonian expectation. We developed an algebraic scheme to calculate
these distribution functions through self-consistent sets of
equations. In this paper, we present an alternative scheme based on
diagrams to calculate these distribution functions. As in the CBF, a
generating functional is introduced and calculated by a linked-cluster
expansion in terms of diagrams. These diagrams are constructed
according to a few simple rules and using only three basic elements:
(a) dots representing the ket-state correlation coefficients, (b)
exchange lines representing the bra-state correlation coefficients,
and (c) Pauli line representing Pauli exclusion principle manifested
by the spin-1/2 operator property $(s^\pm)^2=0$ in our spin model
example. In this fashion, infinite diagrams can be resummed in a
straightforward manner. As a simple application in our spin model, the
spin-wave theory (SWT)\cite{and,tak} is reproduced by including all
so-called ring diagrams without any Pauli line. Approximations beyond
SWT by resummations including Pauli lines are also given. One such
approximation, which includes all so-called super-ring diagrams by a
resummation of infinite Pauli lines in addition to resummations of
ring diagrams, produces a convergent, precise result for the
order-parameter of the one-dimensional isotropic model, contrast to
the well-known divergence of SWT. Furthermore, the diagrammatic
analysis discussed here forms a basis for a possible combination of
the variational CCM and CBF. Such a unified theory may prove to be
capable of dealing with strongly-correlated fermion system in liquid
phase.

\section{The variational coupled-cluster method}

We take the spin-1/2 antiferromagnetic XXZ model on a bipartite
lattice as our example. The Hamiltonian is given by
\begin{equation}
H=\frac 12\sum_{l,\rho }H_{l,l+n}=
\frac 12\sum_{l,n}\left(As_l^zs_{l+n}^z+\frac 12
s_l^{+}s_{l+n}^{-}+\frac 12s_l^{-}s_{l+n}^{+}\right)\;,
\end{equation}
where $A>0$ is the anisotropy constant, the index $l$ runs over all
lattice sites, $n$ runs over all $z$ nearest-neighbour sites, and
$s^\pm$ are the usual spin raising $(+)$ and lowering $(-)$ operators.
As in our earlier work using the traditional CCM \cite{bpx}, we take
the model state as the classical N\'eel state with alternating spin-up
and spin-down sublattices. As before, we shall exclusively use index
$i$ for the spin-up sublattice and the index $j$ for the spin-down
sublattice. The many-spin correlations in its ground state of Eq.~(1)
can then be included by considering the excited states with respect to
the uncorrelated model state. These excited states are constructed by
applying the so-called configuration creation operators $C^\dag_I$ to
the N\'eel model state with the nominal index $I$ labelling these
operators. In our spin model, the operators $C^\dag_I$ are given by
any combination of the spin-flip operators to the N\'eel state $s^-_i$
and $s^+_j$; the index $I$ in this case corresponds to the collection
of the lattice indices ($i$'s and $j$'s). The hermitian conjugate
operators of $C^\dag_I$ are the configuration destruction operator
$C_I$, given by any combination of $s^+_i$ and $s^-_j$. For example,
the two-spin flip creation operator is given by $C^\dag_{ij} = s_i^-
s_j^+/2s$, and their destruction counterpart, $C_{ij}=s_i^+s_j^-/2s$,
where $s$ is the spin quantum number. Although we are mainly
interested in $s=1/2$ in this article, we keep the factor of $1/2s$
for the purpose of comparison with the large-$s$ expansion.

As discussed in details in our earlier paper \cite{yx1}, we use Coester
representation for both the ket and bra ground-states and write
\begin{equation}
|\Psi\rangle=e^S|\Phi\rangle,\quad S=\sum_I F_I C^\dag_I\;;\quad
\langle\tilde\Psi|=\langle\Phi|e^{\tilde S},\quad
 \tilde S=\sum_I \tilde F_I C_I,
\end{equation}
with
\begin{equation}
\sum_I F_I C^\dag_I= \sum_{n=1}^{N/2}
\sum_{i_1...,j_1...}f_{i_1...,j_1...}
\frac{s_{i_1}^{-}...s_{i_n}^{-}s_{j_1}^{+}...s_{j_n}^{+}}{(2s)^n},
\end{equation}
for the ket state and the corresponding hermitian conjugate of Eq.~(3)
for the bra state, using notation $\tilde F_I = \tilde
f_{i_1...,j_1...}$ for the bra-state coefficients. The coefficients
$\{F_I,\tilde F_I\}$ are then determined by the usual variational
equations as
\begin{equation}
\frac{\delta\langle H\rangle}{\delta \tilde F_I} =
\frac{\delta\langle H\rangle}{\delta F_I} = 0\;,
\quad
  \langle H\rangle \equiv 
  \frac{\langle\tilde\Psi| H|\Psi\rangle}
       {\langle\tilde\Psi|\Psi\rangle}\;.
\end{equation}
We define the so-called bare distribution functions as
\begin{equation}
g_I=\langle C_I\rangle, \quad  \tilde g_I=\langle C^\dagger_I\rangle\;,
\end{equation}
where we have exchanged the definition of $g_I$ with that of $\tilde
g_I$ as compared with those in Ref.~7 for purely notational
reason. The Hamiltonian expectation $\langle H\rangle$ is shown, in
general, to be a function containing up to linear terms in $g_I$ and
$\tilde g_I$ and finite order polynomial in $F_I$ (or in $\tilde F_I$)
in Eq.~(21) of Ref.~7:
\begin{equation}
\langle H\rangle = {\cal H}(g_I,\tilde g_I, F_I) = 
{\cal H}(\tilde g_I,g_I, \tilde F_I)\;.
\end{equation}
Two systematic schemes have been developed for calculating the
distribution functions of Eqs.~(5): one is algeraic and the other is
diagrammatic. In the algebraic approach, by taking the advantage of
the properties of the operators, it is straightforward to derive the
following self-consistent sets of equations for the distribution
functions
\begin{equation}
g_I=G(\tilde g_J,F_J)\;,\quad \tilde g_I=G(g_J, \tilde F_J)\;,
\end{equation}
where $G$ is a function containing up to linear terms in $\tilde g_J$
(or $g_J$) and finite order polynomial in $F_J$ (or $\tilde
F_J$). Eqs.~(7) are solved for $g_I$ and $\tilde g_I$ as a function of
$F_I$ and $\tilde F_I$. The variational Eqs.~(4) are then carried to
determined the optimum $F_I$ and $\tilde F_I$. In this algebraic
calculation, direct comparison with the traditional CCM can be
made. It is shown that the CCM is a linear approximation to one set of
distributions as simply $\tilde g_I \approx \tilde F_I$, which is a
poor approximation for the spin-spin correlation function and
low-lying excitations. More detailed comparison was given in
Ref.~7. Here we present the diagrammatic scheme similar to that in CBF
to calculate these distribution functions. We like to point out that
Eqs.~(2) and (4)-(7) are the main general equations of the variational
CCM.

\section{Diagrammatic representation of generating functional}

In this section, we calculate the bare distribution functions $g_I$
and $\tilde g_I$ of Eq.~(5) by employing a diagrammatic scheme. As a
demonstration, we consider a simple truncation approximation in which
the correlation operators $S$ and $\tilde S$ of Eqs.~(2) retain only
the two-spin flip operators as (the so-called SUB2 approximation as
defined in Ref.~10),
\begin{equation}
S\approx \sum_{ij}f_{ij}C^\dag_{ij}
 =\sum_{ij}f_{ij}\frac{s_i^{-}s_j^{+}}{2s}\;,\quad 
\tilde S\approx \sum_{ij}\tilde f_{ij}C_{ij}
 =\sum_{ij}\tilde f_{ij}\frac{s_i^{+}s_j^{-}}{2s}\;. 
\end{equation}
Using the usual angular momentum commutations $[s^z_l,s^\pm_{l'}]=\pm
s^\pm_l\delta_{ll'}$, $[s^+_l,s^-_{l'}]=2s^z_l\delta_{ll'}$, and the
N\'eel state eigenequations, $s_i^z\bigl|\Phi\bigr\rangle =
s\bigl|\Phi\bigr\rangle, s_j^z\bigl|\Phi\bigr\rangle =
-s\bigl|\Phi\bigr\rangle$, it is a straightforward calculation to
derive expectation value of any physical operators in terms
distribution functions of Eqs.~(5). In this approximation, for
example, the expectation value of Eq.~(1) is given by
\begin{equation}
\langle H_{ij}\rangle=A\langle s^z_is^z_j\rangle
 +\frac12(g_{ij}+\tilde g_{ij})\;,
\end{equation}
where $\langle s^z_is^z_j\rangle$ is calculated as
\begin{equation}
\langle s^z_is^z_j\rangle= - s^2+s\left(\sum_{i'}\rho_{i'j}
  +\sum_{j'}\rho_{ij'}\right)
 -\biggl(\sum_{i'j'}\rho_{ij',i'j}+\rho_{ij}\biggr)\;,
\end{equation}
$\rho_{ij}$ is the usual full one-body distribution function defined as
\begin{equation}
\rho_{ij} \equiv f_{ij}\tilde g_{ij}=f_{ij}
\frac{\langle s^-_is^+_j\rangle}{2s}\;,
\end{equation}
and where $\rho_{ij,i'j'}$ is the full two-body distribution function
define as
\begin{equation}
\rho_{ij,i'j'}\equiv f_{ij}f_{i'j'}\tilde g_{ij,i'j'}=
   f_{ij}f_{i'j'}\frac{\langle s^-_is^+_js^-_{i'}s^+_{j'}
         \rangle}{(2s)^2}\;.
\end{equation}
The order parameter is given by
\begin{equation}
\langle s^z_i\rangle = s-\rho\;,
\end{equation}
where $\rho = \sum_j\rho_{ij}$, taking the advantage of translational
invariance.

We define a generating functional $W$ in the usual fashion as,
\begin{equation}
W\equiv \ln \langle \tilde \Psi| \Psi\rangle\;,
\end{equation}
so that the bare and full distribution functions can be simply expressed
as functional derivatives of $W$. For example, the one-body and
two-body bare functions are given by
\begin{equation}
\tilde g_1=\langle C^\dagger_1\rangle =\frac{\delta W}{\delta f_1}\;, \quad
\tilde g_{12} = \langle C^\dagger_1C^\dagger_2\rangle =
 \frac{\delta^2 W}{\delta f_1 \delta f_2}+\tilde g_1\tilde g_2\;,
\end{equation}
where, for simplicity, we have employed notation $1\equiv(i_1,j_1)$ so
that $f_1=f_{i_1j_1}$ etc.; and the structure function $S_{12}$ has
the usual relation as in the CBF as
\begin{equation}
S_{12}\equiv f_1\frac{\delta\rho_2}{\delta f_1} = \rho_1\delta_{12}
  +\rho_{12}-\rho_1\rho_2\;,
\end{equation}
where $\rho_1=\rho_{i_1j_1}$, etc.

We now write $W$ in terms of a linked-cluster expansion
\begin{equation}
W = {\rm sum\;of\;all\;linked\;cluster\;contributions}\;.
\end{equation}
The main task of this section is to find a diagrammatic scheme to
categorize this expansion. We first expand the ket-state operator in
the simplified notation, $e^S =1+S+\frac
1{2!}S^2+\cdots=1+f_1C_1^{\dagger} +\frac
1{2!}f_1f_2C_1^{\dagger}C_2^{\dagger}+\cdots$, where in the last
equation, the summation over all indices is understood. The
normalization integral,
\begin{equation}
\langle \tilde \Psi|\Psi\rangle =1+\tilde
f_{1^{\prime }}f_1\langle C_{1^{\prime }}C_1^{\dagger }\rangle
+\frac 1{(2!)^2}\tilde f_{2^{\prime }}\tilde f_{1^{\prime
}}f_1f_2\langle C_{2^{\prime }}C_{1^{\prime }}C_1^{\dagger
}C_2^{\dagger }\rangle +\cdots\;,
\end{equation}
can be evaluated straightforwardly for the first few terms. In the
above series, the primed indices are used for bra state expansion. We
notice that each term of Eq.~(18) contains equal number of creation
and destruction operators (otherwise, the expectation is zero).

The first-order expectation is easily calculated as $\langle
C_{1^{\prime}}C_1^{\dagger}\rangle
=\frac1{(2s)^2}\langle\Phi|s_{j_{1^{\prime}}}^{-}s_{i_{1^{\prime
}}}^{+} s_{i_1}^{-}s_{j_1}^{+}|\Phi\rangle=\delta _{i_{1^{\prime}}i_1}
\delta _{j_{1^{\prime}}j_1}$. Hence we have, writing out the summation
explicitly,
\begin{equation}
{\rm 1st\; order} = \sum_1f_1\tilde f_1\;.
\end{equation}
The calculation of the second-order expectation $\langle C_{2^{\prime
}}C_{1^{\prime }}C_1^{\dagger }C_2^{\dagger}\rangle$ is slightly more
complicated. We first consider the case of $1\not=2$ (i.e., $i_1\neq
i_2$ and $j_1\neq j_2$). There are four nonzero terms
\begin{eqnarray*}
&&\left( \delta _{i_{1^{\prime }}i_1}\delta _{i_{2^{\prime }}i_2}+\delta
_{i_{1^{\prime }}i_2}\delta _{i_{2^{\prime }}i_1}\right) \left( \delta
_{j_{1^{\prime }}j_1}\delta _{j_{2^{\prime }}j_2}+\delta _{j_{1^{\prime
}}j_2}\delta _{j_{2^{\prime }}j_1}\right) \\
&=&\left( \delta _{i_{1^{\prime }}i_1}\delta _{i_{2^{\prime }}i_2}+\delta
_{i_{1^{\prime }}i_2}\delta _{i_{2^{\prime }}i_1}\right) \left( \delta
_{j_{1^{\prime }}j_1}\delta _{j_{2^{\prime }}j_2}+\delta _{j_{1^{\prime
}}j_2}\delta _{j_{2^{\prime }}j_1}\right) \;.
\end{eqnarray*}
The cases when $i_1=i_2$ and/or $j_1=j_2$ can be easily accounted for
by introducing a factor involving the usual delta functions as,
\[
\left( 1-\frac 1{2s}\delta _{i_1i_2}\right) \left( 1-\frac 1{2s}\delta
_{j_1j_2}\right) =1+\Delta _{12}\;,
\]
with a definition,
\begin{equation}
\Delta _{12}\equiv -\frac 1{\left( 2s\right) }\left( \delta _{i_1i_2}+\delta
_{j_1j_2}\right) +\frac 1{\left( 2s\right) ^2}\delta _{i_1i_2}\delta
_{j_1j_2}\;.
\end{equation}
This is because $\left( s_i^{-}\right) ^2=\left( s_j^{+}\right) ^2=0$
for $s=1/2$, a manifestation of Pauli exclusion principle. The
second-order contribution is hence derived as
\begin{eqnarray*}
&&\frac 1{\left( 2!\right) ^2}\tilde f_{2^{\prime }}\tilde f_{1^{\prime
}}f_1f_2\left( \delta _{i_{1^{\prime }}i_1}\delta _{i_{2^{\prime
}}i_2}+\delta _{i_{1^{\prime }}i_2}\delta _{i_{2^{\prime }}i_1}\right)
\left( \delta _{j_{1^{\prime }}j_1}\delta _{j_{2^{\prime }}j_2}+\delta
_{j_{1^{\prime }}j_2}\delta _{j_{2^{\prime }}j_1}\right) \left( 1+\Delta
_{12}\right) \\
&=&\frac 1{2!}\left[ \left( f_1\tilde f_1\right) \left( f_2\tilde f_2\right)
+f_1f_2\tilde f_{i_1j_2}\tilde f_{i_2j_1}\right] \left( 1+\Delta
_{12}\right) \;,
\end{eqnarray*}
where the second term inside the square brackets clearly represents
the so-called exchange contributions. We notice that we did not
consider explicitly the Pauli exclusion principle for the bra-state
operators in the above derivations as the delta functions for the ket
state operators also take this principle into account due to the fact
that each of the bra-state operators always need to match one of the
ket-state operators in order to give nonzero contribution. This is
also true for higher-order terms.  We also notice the expression of
Eq.~(20) is in fact also correct for spin quantum number $s>1/2$ because,
for a general $s$,
\[
\frac{1}{2(2s)^2}\langle\Phi|(s^+_i)^2(s^-_i)^2|\Phi\rangle 
 =\left(1-\frac{1}{2s}\right)\;,
\]
etc. For higher-order terms in the expansion of Eq.~(18), the
extension of Pauli exclusion principle can be simply written as a
product of two-body factors as
\begin{equation}
\prod_{n>m}\left(1+\Delta_{nm}\right)\;.
\end{equation}
We notice that the above product in general is not exact any more but
an approximation for $s>1/2$ as the three-body effects (e.g., from
$(s^-_{i_1})^3$ when $i_1=i_2=i_3$) have been ignored.

In order to extend to higher-order calculations including the exchange
contributions, we need a systematic graph representation. For this
purpose, as shown in Fig.~1, we use a solid dot to represent the ket
state coefficient $f_1$ with $1=(i_1j_1)$ as defined earlier; a
(directed) exchange line drawing from $i_1$ to $j_2$ to represent the
bra state coefficient $\tilde f_{i_1j_2}$; and a Pauli (dashed) line
drawing between any two dots to represent delta function $\Delta
_{12}$ of Eq.~(20). With these graphic notations, a linked
contribution is represented by a connected diagram. After the detailed
calculations up to 5th order, we have established the following simple
and complete rules for construction of these diagrams in the
normalization integral of Eq.~(18):
\begin{itemize}
\item The $k$th-order contribution consists of all possible diagrams
involving $k$ dots;
\item In each diagram the number of dots equal to number of exchange
lines;
\item A dot is always connected by exchange lines (leaving and coming)
hence exchange lines always form loops;
\item Between any pair of dots one can draw at most one Pauli line;
\item The contribution of each diagram is divided by its symmetry
factor;
\item Summations over all indices involved.
\end{itemize}

We first consider the case without any Pauli line,
$\Delta_{nm}=0$. (This is equivalent to turning spin operators to
boson operators as will be shown later.) For example, the first-order
contribution of Eq.~(19) is simply a dot with an exchange line leaving
and coming as shown as diagram $a$ in Fig.~2, where the direction of
exchange line is clockwise as in most other diagrams (we therefore do
not show arrows of exchange lines explicitly). The second-order
contribution with $\Delta_{12}=0$ is given by two diagrams $b$ and $c$
in Fig. 2, namely $\frac 1{2!}(b+c)$. The 3rd-order contribution is
calculated as $\frac 1{3!}(a+3b+2c)$ and is shown in Fig.~3, where
factor $3$ for diagram $b$ is due to the three equivalent diagrams by
rotation and the factor $2$ for diagram $c$ comes from the two
equivalent diagram with opposite directions, one clockwise the other
counter-clockwise (this is referred to as parity symmetry). In similar
fashion one can write down the 4th-order contribution as shown in
Fig.~4 for the corresponding diagrams as
\begin{equation}
{\rm 4th\; order}=\frac 1{4!}\left( a+6b+8c+3d+6e\right) \;, 
\end{equation}
where the coefficient numbers are the symmetry factors of the
corresponding diagrams. For example, the factor $6$ for diagram $e$ is
due to the fact that there are three equivalent diagrams each with
parity symmetry factor of $2$. The 5th-order contributions include 7
independent diagrams, as shown in Fig.~5, namely
\begin{equation}
{\rm 5th\;order}=\frac 1{5!}\left( a+10b+20c+15d+30e+20f+24g\right)\;. 
\end{equation}
We notice that, in all these results, the last term represents a ring
diagram with $k$ dots in the $k$th order contribution. We use $R_k$ to
represent this ring diagram with the symmetry factor
$(k-1)!/k!=1/k$. For example, the 4th-order ring contribution is,
writing out the summations explicitly
\begin{equation}
R_4=\frac 14\sum_{1,2,3,4}f_1\tilde f_{i_1j_2}f_2\tilde
f_{i_2j_3}f_3\tilde f_{i_3j_4}f_4\tilde f_{i_4j_1}\;. 
\end{equation}
Furthermore, the other terms in these $k$-order contributions are
simply a product of smaller ring contributions. This property can be
extended to higher order. (For this purpose one needs to apply
symmetric group ${\cal S}_n$ to count the number of diagrams. See, for
example, Ref.~11). We are now in position to write all contributions
without any Pauli line in terms of these ring diagrams. The
normalization integral of Eq.~(18) is then written as
\begin{eqnarray*}
\langle\tilde\Psi|\Psi\rangle_{\Delta_{nm}=0}
 &=&\sum_{k=0}^\infty \frac 1{k!}\frac 1{\nu _1!}(R_1)^{\nu_1}
 \frac 1{\nu_2!}(R_2)^{\nu_2}\cdots
 \frac 1{\nu _k!}(R_k)^{\nu _k}\\
 &=&\exp(R_1+R_2+R_3+\cdots)\;.
\end{eqnarray*}
The corresponding generating functional $W'$ without any Pauli line is
simply
\begin{equation}
W'\equiv W\bigg|_{\Delta_{nm}=0} = \sum_{k=1}^\infty R_k\;.
\end{equation}

To include Pauli lines (i.e. $\Delta_{nm}\not=0$), we use notation
$L_k$ to represents the contribution of all linked $k$-clusters and
write
\begin{equation}
W=\ln\langle\tilde\Psi|\Psi\rangle = L_1 +L_2+L_3+\cdots\;.
\end{equation}
Using the simple rules discussed earlier, without much difficulty, we
can list all $k$-cluster contributions of $L_k$ in terms of a ring
diagram $R_k$ plus all possible ways of drawing Pauli lines between
any pair of $k$ dots of rings, including those pairs of dots between
rings and those pairs of dots inside rings. In Fig.~6 we list all
3rd-order contributions in $L_3$ except $R_1,R_2$ and $R_3$.

\section{Diagram resummations, spin-wave theory and beyond}

We first consider all diagrams without any Pauli line, namely all the
ring diagram contributions $R_k$ with $k=1,2,\cdots$, and show that
the spin-wave theory is thus reproduced. As can be seen from Eq.~(20),
these ring diagrams represent the first order approximation in the
large-$s$ limit. In fact, in this limit, operators $s_i^-$ and $s_j^+$
behave like bosons as $s_i^- \rightarrow \sqrt{2s}\;a_i^\dagger$,
$s_j^+\rightarrow \sqrt{2s}\;b_j^\dagger$ \cite{and,tak}. The
corresponding wavefunction by Eq.~(8) becomes the spin-wave function
as
\begin{equation}
|\Psi\rangle \rightarrow |\Psi _{sw}\rangle =\exp
\left( \sum_{ij}f_{ij}a_i^{\dagger }b_j^{\dagger }\right)|\Phi\rangle
 =\prod_q \exp\left(f_qa_q^{\dagger }b_{-q}^{\dagger}\right)|\Phi\rangle\;,
\end{equation}
where the N\'eel state $\left| \Phi \right\rangle $ should be
considered as the vacuum state for the two sets of bosons
$a_i^{\dagger }$ and $b_j^{\dagger }$ and where, in the last equation,
we have made Fourier transformations using the translational symmetry
as,
\begin{eqnarray*}
a_i^{\dagger } &=&\sqrt{\frac 2N}\sum_{\bf q}e^{-i{\bf q\cdot r}_i}
a_q^{\dagger},\quad b_j^{\dagger }=\sqrt{\frac 2N}
\sum_{\bf q}e^{-i{\bf q\cdot r}_j}b_q^{\dagger } \\
f_{ij} &=&\frac 2N\sum_{\bf q}e^{-i{\bf q\cdot }(
 {\bf r}_j-{\bf r}_i)}f_q\;,
\end{eqnarray*}
with summation over $\bf q$ restricted to the magnetic zone. The
normalization integral of Eq.~(27) can be easily calculated as the
wavefunction is uncoupled in $q$-space. Using expansion
$\exp(f_qa^\dagger_qb^\dagger_{-q}) =
\sum_n(f_qa^\dagger_qb^\dagger_{-q})^n/n!$ and a simple algebra
$\langle\Phi|a^n_q(a^\dagger_q)^n|\Phi\rangle = n!$, we have the
following well-known result (see, for example, Chapter~2 of Ref.~12),
\begin{equation}
\langle\Psi_{sw}|\Psi_{sw}\rangle =\prod_q\frac{1}{1-\tilde f_qf_q}\;.
\end{equation}
The corresponding generating functional is hence given by,
\begin{equation}
W_{sw}=\ln\langle\Psi_{sw}|\Psi_{sw}\rangle=
-\sum_q \ln \left(1-\tilde f_qf_q\right)=
\sum_q\left[\tilde f_qf_q +\frac12(\tilde f_qf_q)^2 + \cdots \right]\;,
\end{equation}
which is precisely the result of Eq.~(25) after Fourier
transformation, namely
\begin{equation}
W'=W_{sw}\;.
\end{equation}

Distribution functions without any Pauli line can be easily calculated
using functional derivatives of Eq.~(15)-(16) with diagrammatic
representation. For example, the one-body bare distribution function,
$\tilde g_1' = \delta W'/\delta f_1$, is simply represented by Fig.~7,
where the action of partial derivative is equivalent to unfolding the
ring. Writing out the summations explicitly, we have the expansion of
Fig.~7 as,
\begin{equation}
\tilde g'_1=\tilde f_1+\sum_2 \tilde f_{i_1j_2}f_2\tilde f_{i_2j_1}
 +\sum_{2,3}\tilde f_{i_1j_2}f_2\tilde f_{i_2j_3}f_3\tilde f_{i_3j_1}
 +\cdots\;,
\end{equation}
and similar expansion for $g_1'$. A close inspection of $\tilde g_1'$
and $g_1'$ expansions yields self-consistency equations as
\begin{equation}
\tilde g'_1=\tilde f_1+\sum_2 \tilde f_{i_1j_2}g'_2\tilde f_{i_2j_1}\;,
\quad  g'_1=f_1+\sum_2 f_{i_1j_2}\tilde g'_2 f_{i_2j_1}\;,
\end{equation}
agreed exactly with Eq.~(31) of Ref.~7 in this SWT approximation. The
two-body functions in this approximation can also be easily obtained
in this fashion as given in Ref.~7. The spontaneous magnetization of
Eq.~(13) is given by $\langle s^z_i\rangle = s-\rho'$, with $\rho'$
given by
\begin{equation}
\rho'=\sum_j \rho'_{ij}=\sum_q\frac{\tilde f_q f_q}{1-\tilde f_q f_q}
 = \frac12\sum_q\left(\frac1{\sqrt{1-\gamma_q^2/A^2}}-1\right)\;,
\end{equation}
where we have used the reproduced SWT results of Ref.~7, 
\begin{equation}
\tilde f_q=f_q=\frac{A}{\gamma_q}
(\sqrt{1-\gamma_q^2/A^2}-1)\;,\quad
\gamma_q = \frac{1}{z}\sum_n e^{i{\bf q\cdot r_n}}\;,
\end{equation}
where $z$ is the coordination number and $n$ is the nearest-neighbour
index of the bipartite lattice. For one-dimensional (1D) model at
isotropic point $A=1$, the integral of Eq.~(33) diverges, contrast to
the well-known exact result of $\rho=1/2$ for $s=1/2$ by Bethe ansatz
(see Ref.~10 for references).

To go beyond SWT, we need to include Pauli lines. Using the similar
resummation technique as discussed above, we express the expansion of
bare one-body distribution function $\tilde g_1$ in terms of diagrams
as shown in Fig.~8, similar to the expansion in Chap.~9 of Ref.~12 and
in Ref.~13, after multiplying $f_1$ on both sides of the equation,
\begin{equation}
\rho_1 =f_1\tilde g_1 = f_1 \frac{\delta W}{\delta f_1}
 = {\rm Fig.\;8}\;,
\end{equation}
where we have done all resummations of ring diagrams as in Eq.~(31)
and hence all exchange lines in the diagrams of Fig.~8 are now
function $\tilde g'_{ij}$, not the original exchange line function
$\tilde f_{ij}$. For a simple approximation, we consider the first two
diagrams of Fig.~8 as
\begin{equation}
\rho_1 \approx \rho_1' + \rho_1'\sum_2\Delta_{12}\rho_2'\;.
\end{equation}
We notice that, after ignoring the higher-order $\frac{1}{(2s)^2}$
term, Eq.~(36) (without the common $f_1$ factor) agrees with the
expression $\langle s^+_is^-_j\rangle = \langle s^-_is^+_j\rangle$ of
SWT in Ref.~9 and with Eq.~(31) of our earlier paper Ref.~7 (after
changing the sign of both 5th and 6th terms in the equation as they
were typos). After summing over index $j_1$ with $\rho =
\sum_{j_1}\rho_{i_1j_1}$, we have, using Eq.~(20) for $\Delta_{12}$,
\begin{equation}
\rho = \rho' -\frac{2}{2s}(\rho')^2 +
 \frac{1}{(2s)^2}\sum_j\left(\rho'_{ij}\right)^2\;.
\end{equation}
For $s=1/2$ and isotropic point $A=1$, we obtain $\rho\approx 0.127$
for the square lattice and $0.067$ for the cubic lattice.  They are
smaller than $\rho'=0.197$ and $0.078$ of SWT respectively. This is
not surprising because SWT is known to have over estimated the quantum
fluctuations. The best numerical values for the square lattice vary
from $\rho = 0.16$ to $0.19$, including results from extrapolation of
high-order localized CCM calculations \cite{brx}. For the 1D isotropic
model, however, $\rho$ of Eq.~(37) diverges as $\rho'$ diverges as
mentioned earlier.

We next consider an approximation involving higher-order Pauli lines
by including all higher-order diagrams similar to that of Eq.~(36), as
shown in Fig.~9. This infinite series can again be resummed in a
closed form as a self-consistency equation, equivalent to replacing
$\rho_2'$ in Eq.~(36) by $\rho_2$ itself as
\begin{equation}
\rho_1 = \rho_1' + \rho_1'\sum_2\Delta_{12}\rho_2\;.
\end{equation}
The resummation in Eq.~(38) is similar to the resummation of rings in
Eqs.~(31)-(32), we therefore refer it as super-ring resummation. The
numerical results for $\rho$ thus obtained at the isotropic point for
high dimensions improve slightly, as $\rho=0.145$ for the square
lattice and $0.068$ for the cubic lattice. However, for the 1D
isotropic model, Eq.~(38) produces a convergent, precise number $\rho
= 1/2$. This is interesting indeed, as the divergence of SWT has
troubled theorists for many years. It is worth mentioning that the
traditional CCM SUB2 approximation \cite{bpx2} also produced a
convergent result for the 1D model but at $A=0.373$, not at the
isotropic point $A=1$. We leave more discussion to the following
section, and leave detailed calculations including other higher-order
terms and resummations in two-body function $\rho_{12}$ and structure
function $S_{12}$ of Eq.~(16) somewhere else.

\section{Discussion}

In this article, we present a diagrammatic scheme for the calculations
of distribution functions of the variational CCM, as an alternative to
the algebraic scheme published in our earlier papers \cite{yx1}. The
results of SWT are reproduced by an approximation which resums all
ring diagrams without any Pauli line. Approximations beyond SWT can
also easily be made by including diagrams with Pauli lines. One such
approximation, which includes all super-ring diagrams by a resummation
of infinite Pauli lines in addition to resummations of all ring
diagrams, produces a convergent, precise number for the order
parameter of the 1D isotropic model, contrast to the divergence of
SWT. This cure of SWT divergence is also interesting to 2D models
(including square and triangle lattices) as naive higher-order
calculations within the framework of SWT are also likely to produce
divergent results, despite the fact that the first order results are
reasonable. We believe that similar resummations of super-ring
diagrams as Fig.~9 and Eq.~(38) may provide a solution for such
divergent problems. We leave more detailed calculations to somewhere
else.

It is also possible to include in the ground state higher-order
many-body correlations such as 4-spin-flip operators, in additional to
the 2-spin-flip operators of Eqs.~(8). Furthermore, as demonstrated
here by the diagrammatic approach, a direct link between our
variational CCM and the powerful CBF has now been established, as both
rely on determination of distribution functions through functional
derivatives of a generating functional. In particular, as given by
Eq.~(13), particle density $\rho$ in CBF is equivalent to the order
parameter of our spin models as $\langle s^z_i\rangle = s-\rho$. Its
diagrammatic expansions in two theories are similar (see Chap.~9 of
Ref.~12 and Ref.~13 for more CBF details). For 2D and 3D lattice
models, the values of density $\rho$ are small compared with $s$. Such
spin systems can therefore be described as dilute gases (dilute gases
of quasiparticle magnons of spin waves). For the isotropic 1D model,
density $\rho$ is {\it saturated}, corresponding to the order
parameter equal to zero, a critical value. Our approximation including
a resummation of super-ring diagrams is capable of reproducing
precisely such number. It is also interesting to know that our
diagrammatic analysis of the variational CCM is for the translational
invariance lattice system while similar analysis in CBF is for
inhomogeneous systems \cite{br,ek}.

Clearly, for more accurate results in general, we need to include
correlations between those quasiparticles in our ground state, and the
CBF is well known to be one of most effective theories for dealing
with such particle correlations (even when they are very strong as in
a Helium-4 quantum liquid \cite{fee}) by systematic calculations of
the important two-body distribution functions. We therefore propose a
unified trial wavefunction $|\Psi_U\rangle$ as, including a
generalized Jastrow correlation operator $S^0$ involving quasiparticle
density operator $s^z$,
\begin{equation}
|\Psi_U\rangle = e^{S^0/2}|\Psi\rangle\;,\quad 
 S^0=\sum_{ij}f^0_{ij}s^z_is^z_j\;,
\end{equation}
where $\{f^0_{ij}\}$ are the new additional variational parameters and
$|\Psi\rangle$ is our variational CCM state of Eq.~(2). The
diagrammatic scheme as discussed in this article is useful for
calculating the expansion of the new generating functional of
Eq.~(39). We have made progress in such calculations and wish to
report results soon. We also believe such a unified many-body theory
may prove to be capable of dealing with strongly correlated fermion
systems in general.

\acknowledgments I am grateful to R.F. Bishop for introducing the CCM
to me. Useful discussions with J. Arponen, R.F. Bishop, F. Coester,
and H. K\"ummel are also acknowledged.

\begin{figure}

Fig.~1. Three basic elements for construction of diagrams,
where simplified index notations $1\equiv(i_1,j_1)$ etc. are used.

Fig.~2. First- and Second-order contributions in the expansion
of Eq.~(18).

Fig.~3. Diagrams of third-order contributions in the expansion of
Eq.~(18).

Fig.~4. Similar to Fig.~3 but for the 4th-order contributions.

Fig.~5. Similar to Fig.~3 but for the 5th-order contributions.

Fig.~6. Diagrams of up to third-order contributions to the generating
functional $W$ of Eq.~(26) except ring diagrams $R_1,R_2,R_3$. The
corresponding symmetry factors are, in the same order as the list of
diagrams, $(1/2, 1/2, 1, 1, 1/2, 1, 1/2, 1, 1/3!, 1/2, 1/3)$.

Fig.~7. The ring expansion of the one-body bare distribution
function $\tilde g_1$ of Eq.~(31).

Fig.~8. First few contributions to the full one-body
distribution function $\rho_1$ of Eq.~(35), where open dots indicating
no summations over its indices while solid dots indicating such
summations as before.

Fig.~9. Super-ring diagram expansion, similar to the ring diagram
expansion of Fig.~7 but now involving Pauli lines and with
resummations of ring diagrams already carried out in all exchange
lines. See texts for more details.

\end{figure}

\end{document}